# Brillouin – Mandelstam Light Scattering Spectroscopy:
# Applications in Phononics and Spintronics

Fariborz Kargar[*] and Alexander A. Balandin

Nano-Device Laboratory (NDL) and Phonon Optimized Engineered Materials (POEM) Center, Department of Electrical and Computer Engineering, University of California, Riverside, California 92521 USA

Recent years witnessed much broader use of Brillouin inelastic light scattering spectroscopy for the investigation of phonons and magnons in novel materials, nanostructures, and devices. Driven by developments in instrumentation and the strong need for accurate knowledge of energies of elemental excitations, the Brillouin – Mandelstam spectroscopy is rapidly becoming an essential technique, complementary to the Raman inelastic light scattering spectroscopy. We provide an overview of recent progress in the Brillouin light scattering technique, focusing on the use of this photonic method for the investigation of confined acoustic phonons, phononic metamaterials, magnon propagation and scattering. The Review emphasizes emerging applications of the Brillouin – Mandelstam spectroscopy for phonon engineered structures and spintronic devices and concludes with a perspective for future directions.

[*] Corresponding author (F.K.): fkargar@ece.ucr.edu ; web-site: http://balandingroup.ucr.edu/

Brillouin-Mandelstam light scattering spectroscopy (BMS), also referred to as Brillouin light scattering spectroscopy (BLS), is the inelastic scattering of light by thermally generated or coherently excited elemental excitations such as phonons or magnons. Leon Brillouin and Leonid Mandelstam, the French and the Russian scientists independently predicted and studied interactions between light and thermally excited phonons in solids in early decades of 1900's.[1] However, it was not until 1960's that the experimental BMS research received impetus with the invention of lasers, and, later, introduction of the high-contrast multi-pass tandem Fabry–Pérot (FP) interferometers by Sandercock.[2] Brillouin light scattering can be considered complementary to another inelastic light scattering technique – Raman spectroscopy. In the realm of phonons – quanta of crystal lattice vibrations – Brillouin spectrometer measures energies of acoustic phonons while Raman spectrometer measures energies of optical phonons. In many cases, BMS is a more powerful technique in a sense that it measures not only the energy of phonons near $\Gamma$ point, like Raman spectroscopy, but can provide data for determining the entire phonon dispersion in a large portion of the Brillouin zone.

Owing to the several orders of magnitude smaller energy shifts in the scattered light measured in Brillouin spectroscopy experiments than in Raman spectroscopy experiments, the BMS instrumentation utilizes FP interferometers rather than diffraction gratings. A single plane parallel FP interferometer consists of two flat mirrors launched in parallel configuration with respect to each other, at the spacing of $L$. The wavelength of light, which can be transmitted through the FP interferometer is determined by $\lambda = 2L/m$ (here $m$ is an integer number). In this structure, the free spectral range (FSR) of the interferometer – the frequency difference between two neighboring orders of interference – is $1/(2L)[cm^{-1}]$. The resolution of the instrument is defined by the width of the transmitted peak. The ratio of FSR to the width is referred to as finesse (F). The contrast (C) – the ratio of the maximum to minimum transmission – is a function of the finesse as $C = 1 + 4F^2/\pi^2$, which is less than $10^4$. The low-contrast of the single-plane parallel FP does not allow one to distinguish the low-intensity Brillouin scattering of light from that of the elastically scattered component. The multiphases tandem FP interferometer enhances the spectral contrast by orders of magnitude making it possible to detect the low intensity Brillouin scattering peaks even for opaque materials. The state-of-the-art triple pass tandem FP systems provide a contrast of $10^{15}$ [Ref. 3].

Presently, BMS is at a similar stage of development and use as Raman light scattering spectroscopy was about 30 years ago. The need for determining phonon and magnon dispersions in novel two-dimensional (2D) materials, nanostructures, and spintronic materials resulted in a rapid expansion of this photonic technique to new material systems. Recent developments in the BMS instrumentations, like the confocal Raman spectrometer developments several decades ago, resulted in a much wider use of inelastic Brillouin light scattering. This technique involves acoustic phonons and magnons with hundreds to thousand times lower energy than the energy of optical phonons and magnons, and smaller scattering cross-sections. These differences explain why BMS instrumentation is more complicated than Raman spectroscopy instrumentation and why it took more time to develop.

In the past, Brillouin spectroscopy was mainly used for determining elastic constants of bulk materials and for examining geological samples.[4] More recently, this photonic technique has enjoyed a surge in the use for investigating various physical phenomena in advanced materials and devices. It has been used to investigate the acoustic phonon spectrum changes in phononic metamaterials.[5–20] A demonstration of the acoustic phonon confinement in thin-film membranes, individual semiconductor nanowires (NW), and other nanostructures has been accomplished with this technique.[21–26] Brillouin spectroscopy can measure phonon dispersion in low-dimensional materials and small-size samples where other techniques, *e.g.* neutron diffraction, inelastic X-ray spectroscopy, helium atom scattering, and inelastic ultra-violet scattering (IUVS), fail either due to the sample size or detectable energy range limitations.[27–36] The equipment required for these alternative techniques are also more expansive and voluminous. The accurate knowledge of phonon spectrum is essential for understanding electronic and photonic properties of materials. The application of BMS is not limited to quanta of lattice waves – phonons. Brillouin spectroscopy is proved to be a powerful tool for studying quanta of spin waves (SW) – magnons. The technique was instrumental in demonstrating the condensate of magnons[37] and allowed for *in situ* monitoring of magnon propagation and interaction in spintronic devices.[38–42] BMS technique has been also used in the study of the *noise* of magnons in SW devices.[43] In this Review, we discuss recent developments in physics and engineering enabled with the Brillouin – Mandelstam light scattering spectroscopy, and project the future of this innovative photonic technique.

**Fundamentals of the Brillouin Light Scattering**

It is well known that the light scattering processes depend strongly on the optical properties of materials and the penetration depth of light into the material. As a result, the light scattering must be treated separately for different classes of materials depending on their optical transparency at the excitation laser wavelength.[2,44] In certain cases, the optical selection rules imposed by the size and opacity of material systems under investigation, create new opportunities for BMS technique.[2] Similar to Raman light scattering, which is used for observing optical phonons, BMS is a nondestructive inelastic light scattering of monochromatic narrow-band laser light by acoustic phonons.[2,45] In BMS of crystal lattice vibrations, two different mechanisms contribute to scattering of light: (i) bulk phonons via the volumetric opto-elastic mechanism, and (ii) surface phonons via the surface ripple mechanism.[2,44–46] The dominance of one or the other in BMS spectrum depends on the size, *e.g.* thickness, and optical properties of the material, *e.g.* transparency or opacity of the material at the given excitation wavelength.

***Scattering of light by bulk phonons***: In transparent or semi-transparent materials, where the penetration depth of light is sufficiently large compared to that of the incident laser light wavelength, scattering by the bulk phonons is the dominant mechanism. A propagating acoustic phonon with wave-vector of $\mathbf{q}$ and angular frequency of $\omega_q$ induces local variations in the dielectric constant of the medium which can scatter light through the opto-elastic coupling.[2,45,47,48] A general schematic of the light scattering process by volumetric phonons is illustrated in Fig. 1a. In all light scattering processes, two equations of the conversation of momentum, $\hbar\mathbf{k}_s - \hbar\mathbf{k}_i = \pm\hbar\mathbf{q}$, and the conservation of energy, $\hbar\omega_s - \hbar\omega_i = \pm\hbar\omega_q$, should be satisfied. Here $\mathbf{k}_i$ and $\mathbf{k}_s$ are the wave-vectors and $\omega_i$, and $\omega_s$ are the angular frequencies of the incident and scattered light, respectively. The positive and negative signs on the right-hand side of these equations denote the "anti-Stokes" and "Stokes" processes. In the former a phonon is annihilated (or absorbed) whereas in the latter a phonon is created (or emitted) during the scattering process. The angle between $\mathbf{k}_i$ and $\mathbf{k}_s$, $\phi$, is the scattering angle, which depends on the scattering geometry.[2,49] Since the energy change in the light scattering processes by acoustic phonons is negligible compared to the energy of the incident light, one can assume that $k_i \approx k_s$. Therefore, from the conservation of momentum, it is inferred that the magnitude of the phonon wave-vector is $q = (4\pi n/\lambda)\sin(\phi/2)$, where $n$ and $\lambda$ are the refractive index of the material and the wavelength of the excitation laser light, respectively. In backscattering geometry ($\phi = 180°$), the maximum detectable phonon wave-vector, $q_m = 4\pi n/\lambda$ is achieved.[2]

Apart from the substantial difference in the energies of the elemental excitations, the light scattering processes in both Raman and BMS techniques are similar. The Stokes and anti-Stokes optical peaks appear as doublets on both sides of the central line in the frequency spectrum (Fig. 1b). The spectral analyses of these peaks, including the frequency shift, intensity, and full width at half maximum (FWHM), provides information about the energy, population, and lifetime of the detected elemental excitations.[50] Typical Raman systems can detect the frequency range between ~3 THz (~100 cm$^{-1}$) to 135 THz (4500 cm$^{-1}$). This is a range where the optical phonons reside. The modern low-wave-number Raman (LWNR) instruments utilize novel Notch filters and fine gratings, which allow for detecting phonons with the minimum energies of ~360 GHz (~12 cm$^{-1}$) or higher. The multi-pass tandem FP interferometer optical arrangement deployed in BMS can probe quasiparticles with much lower energies in the range of 300 MHz to 900 GHz.[51] This spectrum interval is essential for probing acoustic phonons and magnons. Both Raman and BMS techniques observe phonons with the wave-vectors close to the Brillouin zone (BZ) center limited by the wave-vector of the excitation wavelength of the visible laser source. A combination of BMS, LWNR, and conventional Raman allows one to detect phonons and magnons in the frequency range from hundreds of MHz up to tens of THz at wave-vectors close to the BZ center.

The dispersion of the fundamental acoustic phonon polarization branches is linear in the vicinity of the BZ center, where the phase velocity ($v_p = \omega/k$) and the group velocity ($v_g = \partial\omega/\partial k$) are equal. Knowing the probing phonon wave-vector and the measured angular frequency from BMS, one can determine $v_p$ and $v_g$ of the observed phonon polarization branches. For example, in the backscattering geometry, which is widely used, $v_p = \lambda f/(2n)$ in which $f$ is the spectral position of the associated peak observed in the BMS spectrum. Since optical phonons have a flat dispersion in the vicinity of the BZ center, determining $q$ is not essential in most Raman experiments. Figure 1c shows a dispersion of the longitudinal (LA) and transverse (TA) acoustic and the longitudinal (LO) and transverse (TO) optical phonons in silicon along [001] direction. Figure 1d provides the actual accumulated Raman and BMS spectra for the same.

***Scattering of light by surface phonons:*** In opaque and semitransparent materials, the BMS spectrum is dominated by the scattering of light by surface phonons via the ripple mechanism.[2,44,45,52,53] The process is illustrated schematically in Fig. 1e,f. Owing to high optical extinction in these types of materials, the penetration depth of light is limited to the surface and only the in-plane component (parallel to the surface in the scattering plane) of the phonon momentum conservation law is satisfied. Therefore, only the phonons with the in-plane wave-vector component, $q_\parallel = k_i \sin(\theta_i) - k_s \sin(\theta_s)$, where $\theta_i$ and $\theta_s$ are the incident and scattered light angle with respect to the normal to the surface, contribute to the light scattering (Fig. 1f). In a complete backscattering geometry where $\theta_i = \theta_s = \theta$, the in-plane phonon momentum is $q_\parallel = (4\pi/\lambda)\sin(\theta)$. Note that the phonon wave-vector depends only on the incident angle and excitation laser wavelength. Under such conditions, one can probe phonons with different wave-vectors by changing the incident angle of the laser light. The latter allows one to obtain the phonon dispersion that is energy (frequency) of the phonons as a function of wave-vector. Obtaining the energy dispersion of the elemental excitations is a distinctive advantage of BMS technique over Raman spectroscopy. Modifications in the phonon dispersion, and correspondingly, in the phonon density of states, contain a wealth of information on confinement and proximity effects in low-dimensional material systems.[54,55]

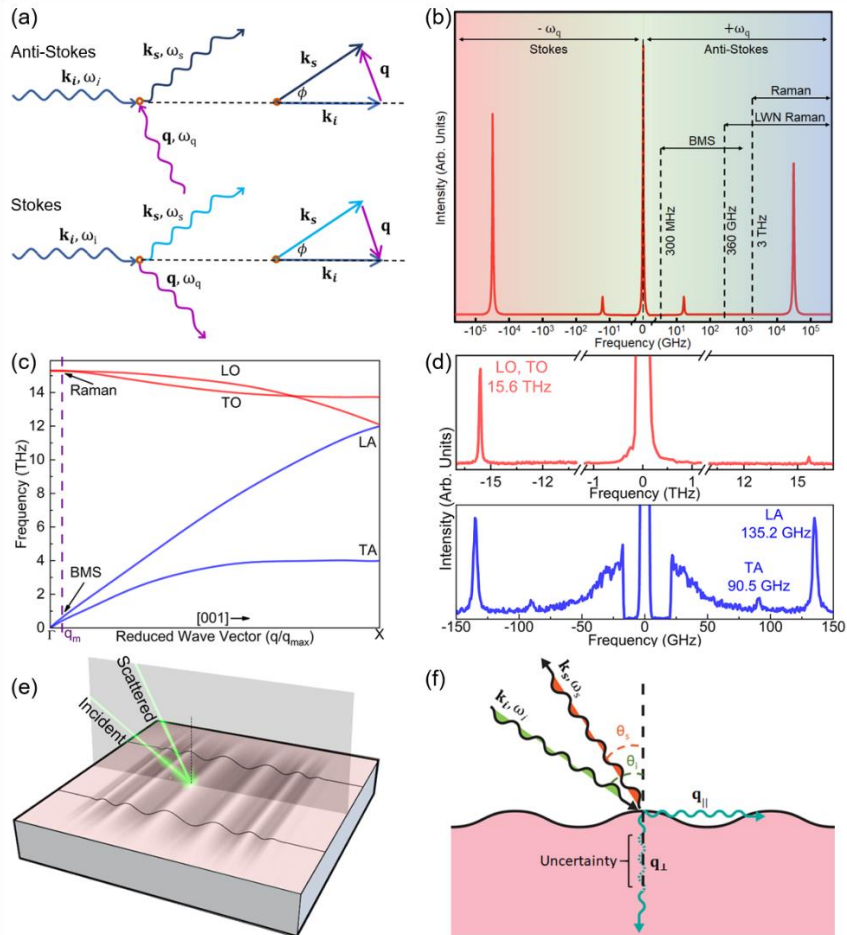

**Figure 1| Fundamentals of Brillouin-Mandelstam light scattering. a)** Schematic of the light scattering processes via bulk quasiparticles. The incident light with wave-vector, $\mathbf{k}_i$, and frequency, $\omega_i$, is scattered to $\mathbf{k}_s, \omega_s$ state either by absorbing (anti-Stokes process) or emitting (Stokes process) a quasiparticle with the wave-vector and energy of $\mathbf{q}, \omega_q$, satisfying the momentum and energy conservation laws**.** Scattering angle, $\phi$, is defined as the angle between $\mathbf{k}_i$ and $\mathbf{k}_s$. **b)** Schematic showing typical spectra and accessible phonon frequency range using Raman, LWNR, and BMS techniques. **c)** Phonon dispersion in silicon crystal along the [001] direction. The dashed line indicates roughly the maximum wave-vector of both optical and acoustic phonons that can be detected by optical techniques. **d)** Raman (up) and BMS (bottom) spectra of silicon showing TO and LO phonons at 15.6 THz and TA and LA phonons at 90.6 GHz and 135.2 GHz at $q = 9.8 \times 10^5$ cm$^{-1}$, respectively. **e)** Schematic showing light scattering by the *surface ripple* mechanism in semitransparent and opaque materials. **f)** Side view of the ripple scattering process where $\theta_i$ and $\theta_s$ are the incident and scattering angles of light with respect to the normal to the surface, $\mathbf{q}_\parallel$ and $\mathbf{q}_\perp$ are the in-plane and normal wave-vector components of the phonons participating in the scattering.

**Phonons in Nanostructured Materials**

Phonons reveal themselves in the thermal, optical, and electronic properties of materials.[55,56] Similar to electron waves, phonon spectrum in nanostructures undergoes changes as a result of either decreasing the physical boundaries to nanoscale dimensions in individual structures[54,57] or as a result of imposing artificial periodicity. [6,58–60] The structures with periodicity, where the phonon dispersion is intentionally modified, are referred to as phononic crystals (PnC).[61] The terminology is similar to the photonic crystals (PtCs) where light propagation in the crystal structure is modified by creating an artificial periodic pattern with a proper period.[62] A new type of structures, termed *phoxonic* crystals (PxC), has been introduced for concurrent modification of both phonon and photon dispersions via artificial periodicity.[63–65] In PxCs, the simultaneous modulation of the elastic and electromagnetic properties is achieved by tuning the material properties such as dielectric constant and mass density, periodic pattern as well as the shape of the individual elements.[63–65] Direct observation of phonon state modifications in these nanostructured materials in the hypersonic frequency range is challenging due to the required high spectral and spatial lateral resolutions. Recent reports demonstrated that BMS is effective technique for observing acoustic phonons in the PnC and PxC samples, which typically have lateral dimensions in the range of a few micrometers.[5–20]

***Phonon confinement in individual nanostructures:*** In the phonon confinement regime, additional phonon polarization branches appear with the optical phonon – like properties such as non-zero energy ($\omega(q=0) \neq 0$) at $\Gamma$ point and nearly flat dispersion at the BZ center.[66–68] These modes are substantially different from the fundamental LA and TA modes, which have zero energies at the $\Gamma$ point. Confined acoustic phonons in different structures share common characteristics as: (i) non-zero energy at the BZ center ($\omega(q = 0) \neq 0$); (ii) quasi-optical dispersion at the vicinity of the BZ center characterized by the large phase velocity but near zero group velocity ($v_g \approx 0$); (iii) decrease in energy difference between different phonon branches with increasing structure dimensions, *e.g.* the nanowire diameter; and (iv) hybrid vibrational displacement profiles, *i.e.* consisting of vibrations along different crystallographic directions.[54] Recent studies reported direct observation of the phonon confinement effects in individual nanostructures such as nanowires,[21,69] thin films,[23,24,70] nanospheres,[22] nanocubes[25], and core-shell structures[71,72] using BMS. In such structures with nanometer-scale dimensions, light scattering by the surface ripple mechanism dominates the spectrum. Figure 2a presents the measured phonon spectrum of a set of long GaAs NWs with diameter of 122 nm and large inter-nanowire distance (red curve) and the reference GaAs substrate without the NWs (blue curve).[21] The inset shows the scanning electron microscopy (SEM) image of the representative sample. The large inter-nanowire distance ensures that the phonon spectrum modification is achieved in the individual nanowires. One can see that the fundamental TA and LA phonon peaks are present in both spectra. Additional peaks are the confined acoustic (CA) phonons in individual NWs. In the spectrum, they are lower in frequency than the bulk TA and LA peaks because the probing phonon wave-vector is different. Figure 2b shows the spectral position of the peaks (symbols) determined using Lorentzian fitting.[21] The data are plotted over the theoretical phonon dispersion calculated from the elasticity equation for an individual NW with the diameter obtained from SEM image analysis. The normalized displacements for a NW at $q = 18.0\ \mu m^{-1}$ and at different frequencies confirm the hybrid nature of these vibrational modes (Fig. 2c). These results demonstrate the power of BMS for obtaining the energy dispersion $E(q)$ or $\omega(q)$ data for individual nanostructures in the samples of small size.

Two decades ago, it has been suggested theoretically that spatial confinement of the acoustic phonons can modify their phase and group velocity, phonon density of states, and the way acoustic phonons interact with other phonons, defects, and electrons.[66–68,73–75] However, direct experimental evidence of such effects was missing. The debated question was also the length scale at which the velocity of phonons undergoes significant changes. A detailed BMS study on silicon membranes with varying thicknesses ($d$) in the range of 7.8 nm to 400 nm confirms that the phase velocity of the fundamental flexural mode experiences a dramatic drop of 15× for membranes with $d \leq 32$ nm.[23] Figure 2d shows the effect of membrane thickness on phonon dispersion as a function of dimensionless wave-vector $q_{\parallel}d$.[23] Note the data enclosed in the dashed box for ultrathin membranes where the phase velocity is significantly lower than the fundamental TA and LA modes as well as the pseudo surface acoustic wave (PSAW) in bulk silicon in the [110] direction. Similar to nanowires and thin membranes, multiple additional phonon branches have been observed in closely packed $SiO_2$ nanospheres with 200 to 340 nm diameters.[22] The acoustic modes are quantized due to the spatial confinement causing the BMS spectrum to be overloaded with many well-defined Lorentzian-shape peaks. The spectral position of these peaks is inversely proportional to the nanosphere's diameter (Fig. 2e). The solid lines in Fig. 2e show the theoretical Lamb spheroidal frequencies of nanospheres, $\nu_{nl}$, where $l = 0,1,2,...$ is the angular momentum quantum number and $n$ is the sequence of eigen modes in increasing order of energy. The BMS selection rules only allow the modes with even numbers of $l$ appear in the spectrum.[76] It should be noted that these peaks are not originated because of the periodicity of nanospheres as their spectral position does not vary with changing the direction of the phonon probing wave-vector in BMS experiment. The direction of the probing wave-vector can be changed simply by in-plane orientation of the sample during BMS experiments. The spectral position of observed peaks plotted as a function of the inverse diameter confirms that the energy of the vibrational eigenmodes decreases with increasing the diameter. Other studies reported the similar confinement effects in GeO$_2$ nanocubes.[25] Up to date, no BMS experiments have been reported to demonstrate the acoustic phonon confinement in quantum dots with diameters in the range of a few nanometers.

The phonon spectrum and related material properties, such as thermal conductivity or electron mobility, can be tuned by embedding nanostructures in materials with a large acoustic impedance mismatch.[54] The acoustic impedance is defined as $\zeta = \rho v_s$, where $\rho$ and $v_s$ are the mass density and sound velocity of each constituents. In this case, the phonon dispersion of the embedded material not only depends on the diameter, material's properties, and surface boundary conditions, but also on the properties of the embedding material, *e.g.* a barrier layer or coating.[77] Interfacing layers of materials with nanometer-scale thicknesses or diameters with a large acoustic impedance mismatch is part of the phonon engineering approach. One can also consider it the *phonon proximity effect* borrowing the terminology from the spintronic and topological insulator fields. Figure 2f present the BMS data for a core-shell structure, where the rigid core silica (SiO$_2$) nanospheres with a diameter of $181 \pm 3$ nm are coated by softer thin layers of polymethyl methacrylate (PMMA) shells with an average thicknesses of 25, 57, and 112 nm.[71] The resulting samples are core-shell particles with the outer diameter ranging from d = 232 nm to 405 nm. For bare silica, two phonon peaks are observed at 13 GHz and 19 GHz. For the core-shell samples, the frequency of these two peaks is suppressed due to the proximity effect of the softer coating layer on the rigid core. As the thickness of the shell increases, the number of vibrational resonance modes in the investigated frequency range also grows (Fig. 2f-bottom). The vibrational profiles of these structures for various *l* are shown in Fig 2g.

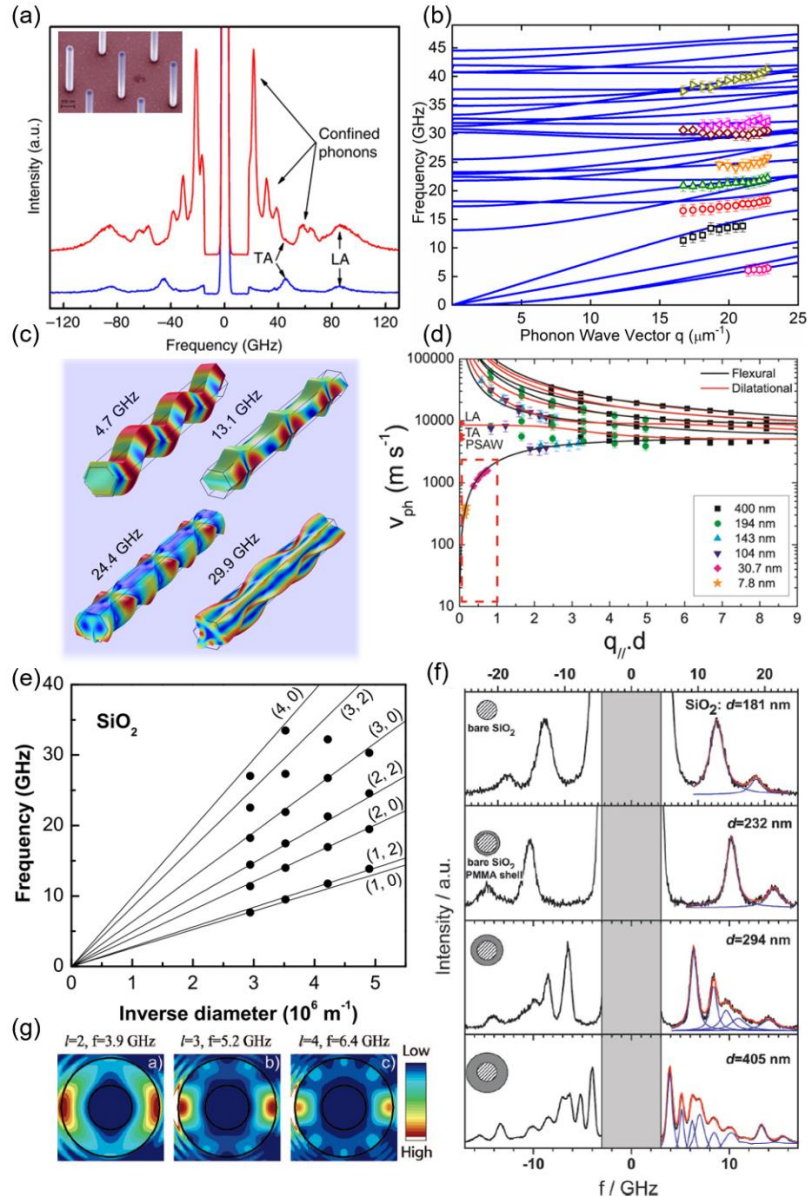

**Figure 2| Observation of phonon confinement in nanostructured materials via BMS technique. a)** Measured BMS spectra of GaAs nanowires with D = 122 nm (red curve) grown on GaAs substrate, and of a pristine substrate (blue curve). Extra peaks in the red curve correspond to the confined acoustic (CA) phonons. The inset shows the pseudo-color SEM image of nanowires. **b)** Measured (symbols) and calculated (blue curves) phonon dispersion of the same GaAs nanowires. **c)** Normalized displacement profiles of confined phonons contributing to Brillouin light scattering. **d)** Phase velocity of the fundamental and confined phonon modes determined from measured BMS data, as a function of the dimensionless wavevector for silicon membranes. **e)** Spectral position of confined phonon peaks (black dots) of silica nanospheres plotted on theoretical frequencies of vibrational eigenmodes denoted by (*n*, *l*) as a function of the inverse diameter. The numbers *n* and *l* denote the sequence of eigenmodes and the angular momentum quantum number, respectively. **f)** BMS spectra of silica nanosphere (top panel) and core-shell silica nanospheres coated with PMMA. Note the difference in the frequency scales of the top and bottom axes. **g)** Displacement profiles of the first three resonance modes of the 405 nm core-shell particle. Panels are adapted with permission from: **a-c:** ref. 21, © 2016 NPG; **d:** ref. 23, © 2012 ACS; **e:** ref. 22, © 2003 APS; **f, g:** ref. 71, © 2008 ACS.

***Phonons in periodic structures:*** As defined above, PnCs are a class of materials consisting of an array of holes or pillars arranged in specific lattice configurations. Figure 3a shows schematics of the square-lattice solid-hole and solid-pillar PnC structures. The periodic modulation of the elastic constants and mass density define the phonon propagation properties. The dimensions of the holes and the period of the artificial lattice define the range of frequencies in which the phonon properties can be engineered.[58] Typically, to modify phonons in the hypersonic frequency range, one needs to fabricate structures with characteristic size of a few tens to hundreds of nanometers.[59] Additional phonon branches appear in the spectrum of PnCs due to localization of phonon modes in its individual constituents or as a result of the Bragg scattering in the periodic structures.[6] The dispersion and energy of these phonon modes can be tailored via changing the characteristic dimensions and the lattice arrangement of the holes or the pillars.

Recent years witnessed an explosive growth of the use of BMS technique for investigation of the acoustic phonon modulation in 1D, 2D, and 3D PnCs.[5–20] Figure 3b shows the measured (solid lines) and calculated (black spheres) phonon polarization branches along the $\Gamma - X$ direction in the square-lattice PnC fabricated on a 250-nm-thick suspended Si membrane.[13] The hole diameters and the lattice constant in this structure are $d = 100\ nm$ and $a = 300\ nm$, respectiveluy. Note that the artificial periodicity causes the BZ edge shrink to $\pi/a \sim 10.47\ \mu m^{-1}$, which is almost three orders of magnitude smaller as compared to that of bulk Si. In this structure, one expects to observe additional phonon polarization branches due to both confined acoustic phonons in the Si membrane as well as phonon folding from the reduced BZ edges owing to the imposed periodicity. Two BMS experiments, one on the pristine Si membrane and another on Si membrane with air holes allow one to assign the associated BMS peaks to either the confined phonons or folded phonons. The phonon folding results in opening a small energy bandgap in the $\Gamma - X$ direction. In this bandgap region, illustrated by the green rectangular in Fig. 3b, propagation of any acoustic waves is prohibited. The pillar-based PnCs allow one to tailor the phonon dispersion by arranging short pillars on top of the bulk substrate or thin membranes. Figure 3c shows the measured and calculated phonon dispersion of a silicon membrane with gold cone-like pillar structure in the square-lattice configuration.[13] The flat dispersion of phonon polarization branches is likely the result of phonon localization in the pillar structures. The vibrational displacement profiles of the phonon branches can be calculated using the elasticity continuum equation. The results of the numerical simulations for a few phonon branches of the air-solid and pillar-based PnCs are presented in Fig. 3d.[13] The red color represents larger displacements. Note the localized modes in the individual pillars for the pillar-based PnCs. The data obtained with BMS is essential for experimental validation of models used for calculation of the phonon dispersion and displacements.

In phoxonic crystals, the phonon and photon states are tailored in the same periodic structure.[64,78–80] Simultaneous tuning of the properties of visible light and acoustic phonons in a specific energy interval requires structures with periodicity in the range of few hundreds of nanometers. A proper design of the lattice geometry and dimensions with contrasting elastic and dielectric properties in PxCs creates an exciting prospect of engineering and enhancing the light-matter interaction.[64,78,79] Such structures can be utilized in designing novel optoelectronic devices such as phoxonic sensors[80–82] and optomechanical cavities.[63,83] Figure 3e shows an SEM image of a designer silicon-on-silicon "pillars with hat" PxCs in the square lattice configuration.[5] The BMS spectra and the Mueller matrix spectroscopic ellipsometry data for phonons and photons are shown in Fig 3f-g, respectively. The BMS measurements along different quasi-crystallographic directions of PxC allow for disentangling the phonon confinement in individual elements of the structure from phonon folding due to the periodicity. The ellipsometry data indicate that the light propagation characteristics are also affected by the structure periodicity. The PxC structure is substantially different from the conventional PtCs, where typically Si pillars are fabricated on a low refractive-index layer, *e.g.* SiO$_2$, to minimize the optical losses.[84]

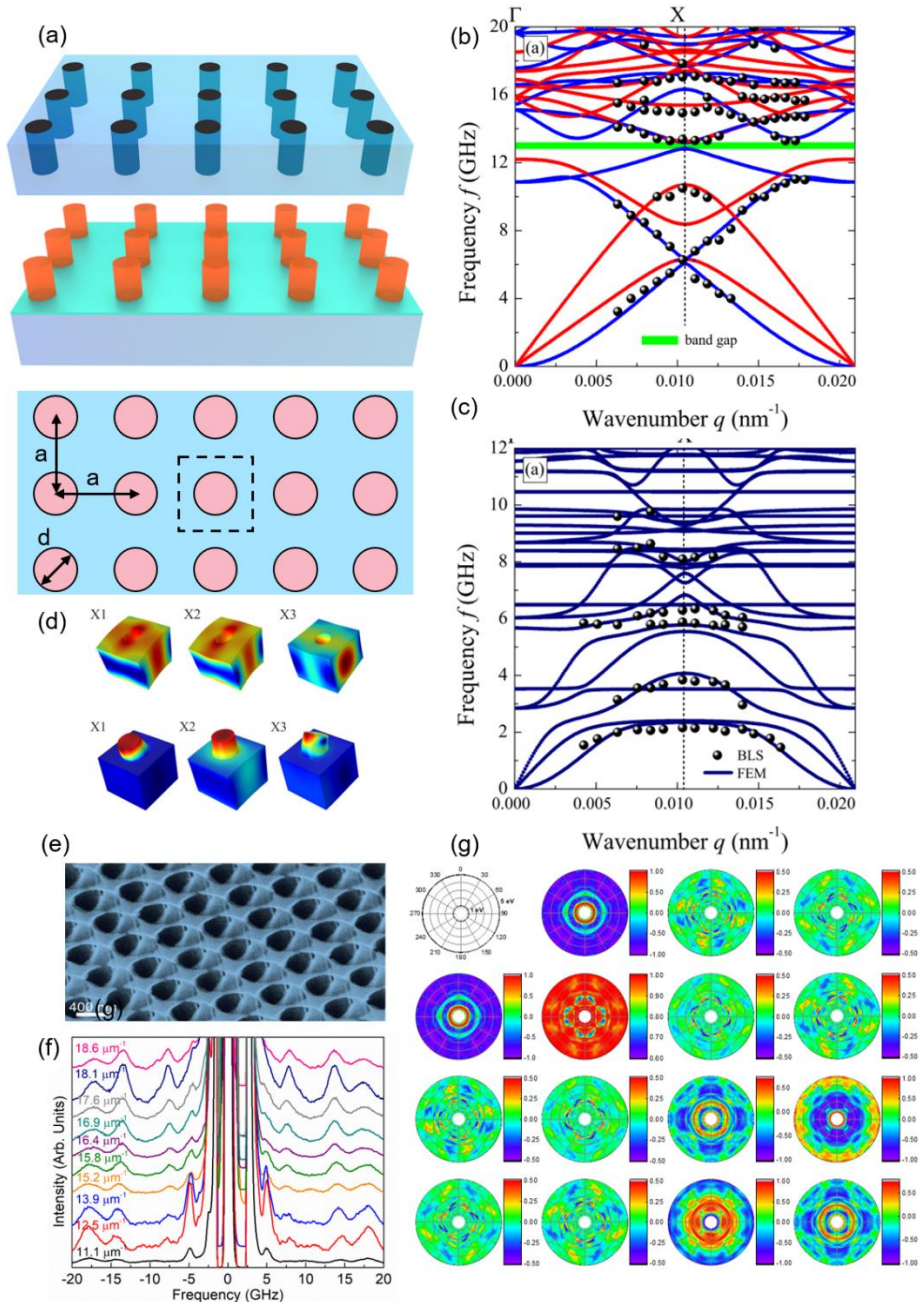

**Figure 3| Phonon spectrum modification in phononic and *phoxonic* crystals investigated by BMS technique. a)** Schematics showing the top and side views of the holey and pillar-based phononic crystals, representing two approaches for engineering the phonon dispersion via artificial periodicity. **b)** Measured (black dots) and calculated phonon dispersion of the holey silicon phononic crystal with the square lattice ($d$ = 100 nm and $a$ = 300 nm) along the $\Gamma - X$ direction. Note the appearance of the phononic band gap depicted with the green rectangular. **c)** Measured (black dots) and calculated phonon dispersion in the Au pillar-based phononic crystal along the $\Gamma - X$ direction. **d)** Displacement profile of the holey and pillar based phononic crystals. **e)** Side-view SEM image of a silicon pillar-based *phoxonic* crystal with the designer shaped "pillars with hats", revealing simultaneous modification of the phononic and photonic properties. **f)** BMS data of the structure shown in **(e)** at different probing phonon wave-vectors. The spectral position of the BMS peaks does not alter with changing the phonon wave-vector confirming the flat dispersion of the phonon branches. **g)** Polar contour plots of the normalized Mueller matrix spectroscopic ellipsometry data at an incident angle of 70° showing 4-fold symmetry for the optical modes in the same structure. Panels are adapted with permission from: **b-d:** ref. 13, © 2015 APS; **e-g:** ref. 5, © 2020 IOP.

**Detection of spin-waves with BMS:** In recent years, BMS has become a standard technique for visualization of SWs and their interactions with other elemental excitations in magnetic materials.[2,85–87] Magnons – quanta of SWs – contribute to light scattering through magneto-optic interaction. The details on how magnons contribute to light scattering have been explained in Ref. 88. The light scattering by bulk magnons follows the same rules of conservation of momentum and energy described above. In the case of surface magnons, such as propagating Damon-Eschbach (DE) modes, the in-plane component of the light scattering process is sufficient to define the magnon's wave-vector.[2]

BMS has several advantages over other experimental techniques such as ferromagnetic resonance (FMR), microwave absorption, or inelastic neutron scattering utilized for detecting SWs.[85] These advantages can be summarized as (i) high sensitivity for detecting weak signals from thermally excited incoherent magnons even in ultra-thin magnetic materials, (ii) space and wave-vector resolution for mapping of SWs, (iii) simultaneous detection of SWs with different frequencies, (iv) wide accessible frequency range, which became possible with FP interferometry, and (v) its compactness of instrumentation.[85] These features of BMS made it essential for recent progress in the magnonic research, which include accurate spatial mapping of externally excited SWs, observation of the Bose – Einstein magnon condensation, and investigation of the Dzyaloshinskii – Moriya interaction (DMI).[39,40,43,89–115] In all these examples of the use of BMS, it was important to have precise positioning of the excitation laser beam on magnetic samples with micrometer-scale lateral dimensions and keep it stable over the long data accumulation times.[85] In the modern-day instrumentation, these requirements are satisfied in the fully automated micro-BMS (μ-BMS), which can monitor and compensate the positional displacement of the sample due to temperature drifts. [85,113] Below, we describe the recent breakthroughs achieved in spintronic – magnonic field with μ − BMS systems in more details.

Magnon currents or SWs can be externally excited in magnetic waveguides using antennas and microwave currents. A number of studies have been devoted to investigating the SW transport in waveguides implemented with ferrimagnetic insulators, *e.g.* yttrium-iron-garnet (YIG), ferromagnetic materials such as permalloy (Py), or other material systems.[38,90–92,116,117] By spatially mapping the intensity of the magnon peak in Brillouin spectrum as a function of the distance from the emitting antenna, one can determine the SW decay length or damping parameter.[90–92,114–116,118] The high sensitivity of BMS allows for detection of the magnon peaks even at millimeter-scale distances from the emitting antenna.[42] The BMS detects nonlinear effects such as second order SWs in the multi-magnon scattering processes.[117,118] Figure 4a shows an optical microscopy image of a two-dimensional Y-shaped SW multiplexer in which the SW dispersion and propagation in Py can be controlled by the local magnetic fields induced by the current applied to each Au conduit.[92] The SWs are launched into the structure by a microwave antenna in the frequency range between 2 GHz to 4 GHz, and are routed to either left or right arm via passing the DC current by connecting either the S1 or S2 switch, respectively. The BMS peak intensity as a function of the excitation frequency in each arm of the Y-shaped structure is presented in Fig. 4b. The spectra confirm a possibility of efficient SW switching in this waveguide design. Figure 4c presents a two-dimensional BMS intensity mapping of SW propagation at 2.75 GHz excitation, which clarifies that SW travels in the same direction as the current flow.

A reported possibility of the magnon Bose-Einstein condensate (BEC) at room temperature (RT) in magnetic materials is one of the most intriguing findings demonstrated with $\mu-$BMS technique.[37] It has been postulated that BEC is achieved if magnons density exceeds a critical value by either decreasing the temperature or increasing the external excitation of magnons.[37] Previously, BEC was demonstrated at low temperatures.[119,120] The condensation can occur at relatively high temperatures if the flow rate of the energy pumped into the system surpasses a critical threshold.[37] In the first BEC demonstration with Brillouin scattering,[37] magnons were excited in a YIG film by external microwave parametric pumping field with a frequency of $2\nu_p$ such that $\nu_p > \nu_m$ in which $\nu_m$ is the minimum allowable frequency of the magnon dispersion at the uniform static magnetic field.[37] The pulse width of the pumping ranged from 1 μs to 100 μs. The microwave photon with a frequency of $2\nu_p$ creates two primary excited magnons with a frequency of $\nu_p$ with the opposite wave-vectors. The intensity and frequency of magnon peaks were recorded using the time-resolved BMS at the delay times after pumping in the range of hundreds of nanoseconds. The scattering intensity, $I_\nu$, at a specific frequency is proportional to the reduced spectral density of magnons, which is proportional to the occupation function of magnons, $n_\nu$. The utilization of an objective with large numerical aperture (NA) allowed to capture magnon modes with a wide range of wave-vectors. The growing intensity, decreasing frequency of the magnon modes from $\nu_p$ to $\nu_m$ in larger delay times, and spontaneous narrowing of the population function[37,121] indicated that the excited magnons were condensed to the minimum valley of the magnon band at RT.

These experimental results led to intensive developments in the field. One theoretical study argued that in the implemented scheme the magnon condensate would collapse owing to the attractive inter-magnon interactions.[122] However, another experimental study demonstrated that the interaction between magnon in condensate state is repulsive leading to the stability of the magnon condensate. The schematic of the experiment is shown in Fig. 4d.[123] The cross-section of the experimental setup is shown in Fig. 4e. The dielectric resonator at the bottom parametrically excites the primary magnons in the YIG film similar to the previous study.[37] The DC current in the control line, placed between the resonator and the YIG film, creates a local non-uniform magnetic field, $\Delta H$, which adds to the static uniform magnetic field, $H_0$. The spatial distribution of the condensate magnons are probed by μ-BMS along the magnetic field by focusing the laser beam on the YIG film surface. The local variation of the magnetic field creates either a potential well or a potential hill depending on its orientation, which adds to the uniform static magnetic field (Fig. 4f.). Figure 4g shows the recorded BMS intensity representing the condensate density along the *"z"* direction.[123] These results are obtained under the stationary-regime experiments where both the pumping and inhomogeneous field are applied continuously. They show that in the case of potential well ($\Delta H_{max} = -10$ Oe), the maximum condensate density occurs in the middle of the control line and it reduces significantly outside the potential well. In the case of potential hill, $\Delta H_{max} = +10$ Oe, an opposite behavior is observed where the density of the condensed magnons shrinks at the center, and it gradually increases towards the outside of the hill. The latter suggests that the condensed magnons tend to leave the area of the increased field resulting in minimum condensed density at the center. This behavior contradicts the assumption of the attractive inter-magnon interaction and necessitates a repulsive interaction among the condensed magnons.[123]

It is known that μ-BMS is one of very few techniques that can be used to investigate the DMI strength in magnetic materials and heterostructures.[124–128] DMI is the short-range antisymmetric exchange interaction in material systems lacking the space inversion symmetry.[124–128] It leads to non-reciprocal propagation of SW, thus providing a way to quantify its strength. Brillouin spectrum typically consists of the frequency-wise symmetric phonon and magnon peaks from Stokes and anti-Stokes processes (Fig 1b). Due to the asymmetry induced by DMI in the dispersion of the SWs, a small detectable frequency difference occurs between the Stokes and anti-Stokes peaks with the wave-vectors of $(-q)$ and $q$, respectively. Without the DMI, the SW dispersion is frequency-wise symmetric and there would be no energy difference between SWs with $q$ and $-q$ wave-vectors. This is shown schematically in Fig. 4h where the dispersion of the SW in the absence of DMI (dashed) and with DMI (solid curves) for $q \parallel \pm x$ and in-plane magnetization of $M \parallel \pm z$, respectively.[93] The high sensitivity of the BMS technique allows detection of the even small frequency difference caused by weak DMI interaction. The bottom panel of Fig. 4h shows the actual BMS data for a heterostructure of 1.3 nm thick Py ferromagnetic layer on a 6 nm thick high spin-orbit heavy metal Pt at fixed $q = 16.7 \ \mu m^{-1}$ under external magnetic field of $\pm 295$ mT.[93] Note the frequency difference of the Stokes and anti-Stokes peaks' spectral positions which is ~0.25 GHz.

**Magnon spatial confinement effects and magnonic crystals:** Similar to phonons, magnon dispersion can be modified due to the size effects in individual structures[51,87,129–133] or in the periodic magnetic structures referred as *magnonic* crystals.[85,134] Several studies reported modifications in the magnon energy dispersion in such structures using the BMS technique. [51,85,87,129–134] The dispersion of magnons can also be tuned by external stimuli induced via strain.[51] Figure 4i shows a structure in which a thin polycrystalline layer of Ni, a magneto-strictive material, is deposited on a PMN-PT ($[Pb(Mg_{1/3}Nb_{2/3})O_3]_{(1-x)}$–$[PbTiO_3]_x$) piezoelectric substrate. [51] The finite thickness of the Ni layer results in quantization of the magnon states giving rise to perpendicular standing spin waves (PSSW) across the Ni layer. By applying a DC voltage to the piezoelectric substrate, a strain-field is induced, and the energy of PSSW modes downshifts owing to the magneto-elastic coupling effect (Fig 4i, bottom panel). In this BMS study, the non-monotonic dependence of the PSSW modes is attributed to difference in the pinning parameters at Ni-air and Ni-substrate interfaces.

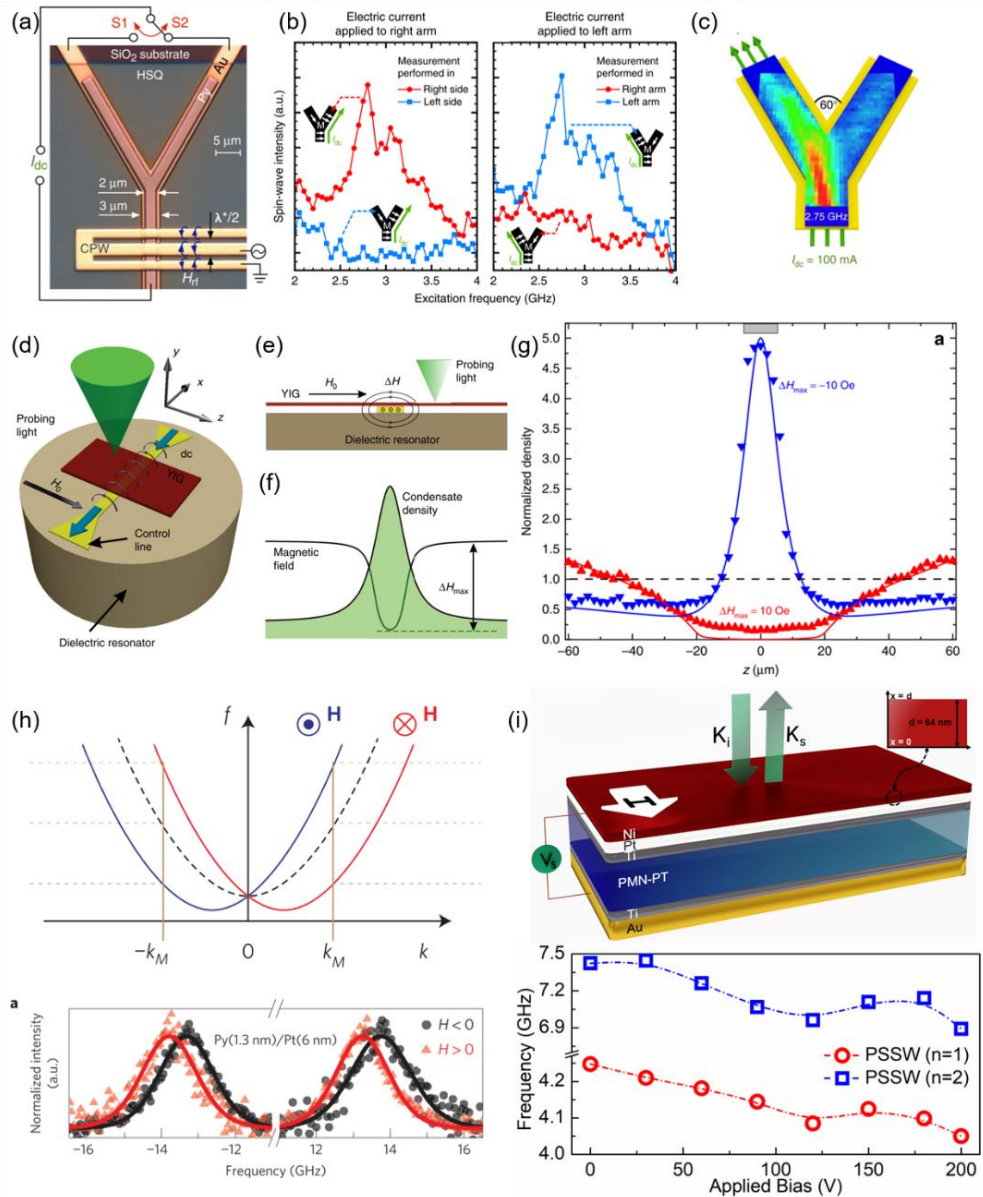

**Figure 4| Investigation of spin wave phenomena and magnon transport using BMS. a)** Optical image of a device structure for investigation of spin wave propagation in the Y-shaped Py waveguide. **b)** Measured BMS intensity for various excitation frequencies of propagating spin waves inside the left (blue curve) and right (red curve) arms at 4.5 µm distance from the Y junction. **c)** Contour map of the BMS intensity of the propagating spin-waves. **d)** Schematic of the Bose-Einstein magnon condensation experiment where magnons in the YIG film are excited by the microwave-frequency magnetic field created by a dielectric resonator. **e)** Side view of the same setup shown in (**d**) exhibiting the external field, $H_0$, and the magnetic field created by the resonator. **f)** Spatial distribution of the horizontal component of the magnetic field, $H_0 + \Delta H$, and magnon condensate density created by the inhomogeneity of the field. **g)** The normalized condensate density recorded by BMS for two cases of the potential well (blue) and potential hill (red). **h)** Schematic of the BMS spectrum with (solid) and without (dashed) the interfacial DM interaction. Bottom panel shows the asymmetry in the spectral position of Stokes and anti-Stokes peaks of the Damon-Eschbach spin waves in Py/Pt as a result of DM interaction. **i)** (up) Device structure with FM polycrystalline nickel thin-film layer deposited on the PMN-PT piezoelectric substrate. With applying the bias to PMN-PT, a biaxial strain is induced in the upper nickel layer, which affects the frequencies of PSSW modes (bottom). Panels are adapted with permission from: **a-c**: ref. 92, © 2014 NPG; **d - g**: ref. 123, © 2020 NPG; **h**: ref. 93, © 2018 APS; **i**: ref. 51, © 2020 Elsevier.

## Outlook

In recent years, BMS has proven itself as a versatile nondestructive photonic technique for applications in solid-state physics and engineering research. The capabilities offered by BMS have already resulted in advancements in the fields of low-dimensional magnetic[85–87,134] and non-magnetic materials and nanostructures,[5,6,21,24,52,135–137] polymers,[12,138–148] biological systems,[149–154] and imaging microscopy.[152,155–158] One can foresee that this technique will find even broader use in investigations where handling the small-size samples and detecting elemental excitations with small energies are essential. The perspective future research directions for BMS include, but not limited to, observation of topological and protected phonon states in phononic metamaterials,[159–163] phonon chirality,[164–166] observation of phonons in hydrodynamic regime,[167–172] investigation of interaction of elemental excitations in bulk and low-dimensional magnetic and ferroelectric materials.[39,50,94,173] BMS is promising for studying the theoretically predicted topologically protected phonon states and one-way acoustic wave propagating modes in phononic metamaterials.[159–163] BMS can provide the full dispersion of hypersonic phonon modes, in GHz frequency range, through the complete first and higher order BZs.[5,6] The artificial periodicity of the phononic metamaterials shrinks BZ to the accessible range of wave-vectors detectable by BMS. It is expected that once the obstacles with fabrication of such complicated material systems with topological-dependent properties are addressed, BMS would become preferential experimental approach since other non-optical methods would fail either due to the sample size limitations or complicated nanofabrication procedures required for other types of measurements.

One can envision a broader use of BMS in the study of phonons in graphene and other quasi-2D

and quasi-1D van der Waals materials. Despite more than a decade of investigation of phonon thermal transport in graphene, there are many open questions. For example, the Grüneisen parameters, and even velocities of the acoustic phonon modes, which carry heat in graphene, have not been accurately measured yet. The problem is that the conventional BMS spectrometers have been limited by inability of locating the samples with lateral dimensions smaller than few micrometers as well as the reduced light scattering cross-section in the low-dimensional materials. Moreover, detection of elemental excitations with frequencies lower than ~1 GHz is challenging. At the phonon wave-vectors of interest, the out-of-plane TA phonon frequencies in graphene and many other 2D materials are lower than the cut-off frequency. The state-of-the-art BMS systems are capable of measuring frequencies down to ~300 MHz, which is important for observation of the out-of-plane (ZA) acoustic phonons in graphene and other low-dimensional materials. Technically, the minimum accessible frequency is limited by the excitation laser's linewidth which is ~100 MHz for BMS applications. The small scattering cross-section for many light scattering processes in low-dimensional materials require long data accumulation times. The modern BMS equipped with additional anti-vibrational systems can overcome this hurdle and can be run for days of measurements. Phonon chirality is another interesting concept in low-dimensional materials, which has been experimentally demonstrated via indirect measurements.[164–166] It is anticipated that BMS can provide a direct observation in certain material systems with trailed dimensions and structures. Another use for BMS technique can be derived from a classical theoretical study, which suggests that Brillouin spectroscopy would be a suitable technique for investigating phonons in the hydrodynamic regime, where the macroscopic collective phonon transport occurs, which is neither ballistic nor diffusive.[167] Recent studies predicted that owing to the modification of phonon states in the low-dimensional materials, the hydrodynamic phenomenon can happen at substantially higher temperatures than previously believed.[168,171]

The interaction between elemental excitations, such as phonon-magnon coupling in magnetic materials is another field attracting a lot of attention in recent years. Although the first attempts of studying such interactions by BMS dates back to three decades ago in bulk YIG,[174] it has been reignited by the discovery of low-dimensional magnetic and anti-ferromagnetic materials.[175–178] BMS has been used for measuring local temperature in the studies related to spin caloritronic, where the interplay between the spin and heat transport is of interest.[50,133] It appears that the BMS technique can follow the same expansion of the use trajectory as Raman spectroscopy and Raman optothermal technique.[179,180] Recent examples of new BMS designs, *e.g.* BMS systems with the beyond optical diffraction limit resolution[115] and rotating microscopy as well as the use of AI capabilities for probing samples with lateral dimensions below the micrometer scale, will elevate this photonic technique to absolutely new level of capabilities.

**Acknowledgements**

AAB and FK acknowledge the support of the National Science Foundation (NSF) via a project Major Research Instrumentation (MRI) DMR 2019056 entitled Development of a Cryogenic Integrated Micro-Raman-Brillouin-Mandelstam Spectrometer. AAB also acknowledges the support of the Program Designing Materials to Revolutionize and Engineer our Future (DMREF) via a project DMR-1921958 entitled Collaborative Research: Data Driven Discovery of Synthesis Pathways and Distinguishing Electronic Phenomena of 1D van der Waals Bonded Solids; and the support of the U.S. Department of Energy (DOE) via a project DE-SC0021020 entitled Physical Mechanisms and Electric-Bias Control of Phase Transitions in Quasi-2D Charge-Density-Wave

Quantum Materials. The authors thank Zahra Barani for her help with preparation of schematics in Figure 3 (a-b).